\documentclass[journal=jacsat,manuscript=article]{achemso}

\usepackage[version=3]{mhchem} 

\author{Bar Ezra}
\email{bar.ezra@mail.huji.ac.il}
\affiliation[The Hebrew University of Jerusalem]
{The Institute of Chemistry and the Fritz Haber Centre for Theoretical Chemistry, The Hebrew University of Jerusalem, Jerusalem 9190401, Israel}
\author{Shimshon Kallush}
\email{shimshonk@hit.ac.il}
\affiliation[The Hebrew University of Jerusalem]
{The Institute of Chemistry and the Fritz Haber Centre for Theoretical Chemistry, The Hebrew University of Jerusalem, Jerusalem 9190401, Israel}
\alsoaffiliation[HIT Holon Institute of Technology]
{Sciences Department, Holon Academic Institute of Technology, Holon 58102, Israel}
\author{Ronnie Kosloff}
\email{ronnie@fh.huji.ac.il}
\affiliation[The Hebrew University of Jerusalem]
{The Institute of Chemistry and the Fritz Haber Centre for Theoretical Chemistry, The Hebrew University of Jerusalem, Jerusalem 9190401, Israel}

\title{Simulating photo-dissociation in strong field enabled by the random phase thermal wavefunction approach}

\abbreviations{}
\keywords{Photodissociation, Velocity Mapping, Quantum Simulations, Random phase thermal wavefunction}

\usepackage{bm}
\usepackage{braket}
\usepackage{graphicx}
\usepackage{subcaption}
\usepackage{wrapfig}
\usepackage{svg}

\usepackage[utf8]{inputenc}
\usepackage[T1]{fontenc}
\usepackage{mathptmx}
\usepackage{multicol}

\usepackage{xcolor}

\begin{document}

\begin{abstract}
    Simulating photo-dissociation processes is a challenging task when the number of states involved becomes significantly large. Typically, the computation complexity is reduced by enforcing restrictions that limit the scope of the model. We present an ab-initio quantum model for strong field photo-dissociation processes which incorporates explicitly rotational dynamics. The large computational complexity was reduced by employing the random phase thermal wavefunction method. The simulation outcome are analogous to experimental observables, such as the momentum angular distribution of the photo-fragments. The present work focuses on studying the convergence of such observables at two different field intensities. The simulation method can be applied to wide-ranging time-domain spectroscopy at experimental conditions far beyond the reach of accurate exact numerical methods. In addition to the conventional dependency on the system's size and the number of realizations, the convergence of the method is found to depend on the level of coherence of the simulated observable.
\end{abstract}

\section{Introduction}

Time domain spectroscopy can be directly modeled by solving the time-dependent Schrödinger equation. Direct modeling for a strong field photo-dissociation process has almost no alternative, since analytical methods are not applicable and perturbation expansions do not converge. This work presents an a first principle simulation method applicable to the photo-dissociation process of a diatomic molecule. The simulation is guided by strong field pulsed experiments. The goal is that the simulation outcome could be directly comparable to experimental observables, specifically to the velocity distribution of the photo-fragments. We define an ab initio simulation to be free of empirical parameters: based on the Hamiltonian, initial temperature, charge, and mass.  

The knowledge of the total Hamiltonian and initial state is required to solve the time-dependent Schrödinger equation. The Born–Oppenheimer decomposition allows to separate the electronic and the nuclear terms\cite{born1927quantentheorie}. Ab initio electronic structure methods are able to accurately calculate the electronic potential surfaces and the transition dipole moments associated with the photo-dissociation. The present study concentrates on solving the nuclear dynamics, specifically the dynamics of the vibrational and rotational degrees of freedom of a diatomic molecule
on multiple electronic surfaces. The strong coupling between the molecular degrees of freedom opens multiple channels to unbound electronic states where the molecule dissociates. 

Within the common experimental framework \cite{shahi2017intense,shahi2019ultrafast}, various physical measures of the photo-fragments are observed far from the dissociation point. The simulation should be able to calculate the same observable, and specifically the angular distribution of the outgoing momentum. 
Such a calculation is set in a large phase-space volume, which translates to a very large computation Hilbert space. In the weak field regime one can use perturbation theory, and the calculation is simplified significantly since selection rules limit the accessible Hilbert space. In strong field  photo-dissociation processes, many interference channels become open. As a result, the calculation has to include all the relevant quantum states that participate in the dynamical process.

Solving the time-dependent Schrödinger equation for a large Hilbert space is a difficult task. The computation effort scales with the size of the Hilbert space, the product of the energy range, and time period.
The Hilbert space size is determined by the phase-space volume containing the problem. The phase-space is restricted by the maximum energy the molecule can carry and by the asymptotic distance estimated by the effective dissociation point.
The number of angular momentum states required for the calculation is determent by the initial temperature of the molecule and the strength of the radiation field. Each participating photon changes the rotational angular momentum by one unit of $\hbar$. Therefore, when the molecule is subjected to a strong radiation field, the size of the manifold of the accessible states increases significantly. For a photo-dissociation study the Hilbert space is vastly, of the order $10^6$ states for the representation of the whole ro-vibronic manifold (where, the vibration is represented on a grid of size ($\sim10^3$), rotational states $(\sim10^2)$ and spin-orbital states $(\sim10^1)$)\color{black}. In principle, the initial state is a thermal mixed state formally requiring  a density operator description. The dimension of the density operator is the square of the Hilbert space, this means that such a calculation is prohibitively expensive. 

Several studies have addressed the computational complexity of similar systems. In the study of $I_2^-$, one of the most studied diatomic molecule, the complexity is reduced by neglecting the rotational degree of freedom of the heavy molecule  \cite{hoops2003ExpI2,davis2003timeI2,Zanni1999femtosecondI2}. For the $H_2^-$ case, the lightest molecule, the study was limited to low initial temperatures, resulting in only the ground ro-vibrational state \cite{wang2006dissociationH2,anis2008roleH2,mckenna2012controllingH2}. A commonly employed approximation assumes that the initial orientation is fixed to the perpendicular direction of the polarization of the laser field. This leads to a significant reduction of the accessible Hilbert space (practically to $M=0$ only) \cite{sun2019mappingK2}
with the price that the simulation scope is limited.

The random phase wavefunction (RPW) method is employed in this study to overcome the challenge of the large Hilbert space. 
It replaces a density operator calculations by a finite sample of random phase thermal wavefunctions
\cite{gelman2003simulating}.
Previously the method has been used to calculate the dynamics of molecular systems including rotational states \cite{kallush2015orientation,hu2021investigation,HU2021139052}. The present study is aimed to develop methods to calculate the full asymptotic photo-fragment distribution. This requires to study the convergence of the calculation with respect to the number of RPW employed.
Typically in the RPW theory the convergence of the observables scales with the number of realizations, $K$, and the Hilbert size, $N_S$, as  $\sim \frac{1}{\sqrt{K}}\cdot \frac{1}{\sqrt{N_S}}$ \cite{kallush2015orientation}. Our purpose is to investigate the prefactor of this scaling, or particularly what is the number $K$ of RPW that is required to converge  a specific observable to experimental accuracy. 

The developed simulation is general, it can be used to calculate a variety of processes and systems. We choose to model the $F_2^-$ system recently studied experimentally by Strasser \textit{et al.} \cite{shahi2017intense,shahi2019ultrafast} . For this system accurate ab-initio potentials have been calculated (private communication with Prof. Moszynski ).
Therefore if the full nuclear dynamical calculation can be converged, the $F_2^-$ system can be simulated from first principles. {Note however, that the goal of the present work is to study the random phase method and for this task we simplified the system and employed only two of the four potentials. (see below) }\color{black}

\section{Methods}
\label{sec:model}

The dynamics of a gas phase diatomic molecule takes place on $n$ electronic surfaces and three ro-vibrational internuclear degrees of freedom, the internuclear distance, and the two rotational angles. The system Hamiltonian includes the nuclear Hamiltonian of all electronic states, and the radiation coupling elements between the states are mediated by the transition dipole moment. In the Born–Oppenheimer decomposition, \cite{born1927quantentheorie} the Hamiltonian can be written as:

\begin{eqnarray}
    \hat{H}_{sys} = \sum_n \left( \hat{H}_n\otimes\left|n\right>\left<n\right| + \sum_{k\neq n} \boldsymbol{\mu}_{n,k}\cdot\boldsymbol{\varepsilon}\otimes\left|n\right>\left<k\right|\right)
    \label{eq:Hamiltonian}\\
    \hat{H}_n = \frac{\hat{P}_{r}^2}{2m_r}+V_n\left(r\right)+\frac{1}{2I}{\hat{\textbf{R}}}^{2}\left(\theta,\phi\right)
\end{eqnarray}
where $\hat{H}_n$ is the nuclear Hamiltonian operator of the $n^{th}$ electronic state. {$\hat{H}_n$ contains }\color{black} the electronic potential of the $n$ electronic state, $V_n\left(r\right)$, {at each internuclear distance $r$,} \color{black} and the total nuclear kinetic energy. The kinetic energy is composed by the linear motion term, $\frac{\hat{P}_{r}^2}{2m_r}$, and the rotational motion term,$\frac{1}{2I}{\hat{\textbf{R}}}^{2}\left(\theta,\phi\right)$. Here, { $P_r$ is the linear momentum}\color{black}, $I(r)=m_rr^2$ is the moment of inertia, and $\hat{\textbf{R}}\left(\theta,\phi\right)$ is the nuclear rotational angular momentum operator. The laser field, $\varepsilon$, is a transform limited pulse with the parameters presented in table \ref{Tal.Para}.   \par

\begin{table}[h!]

\centering
\caption{Model parameteres}\label{Tal.Para}
\label{table-1}
\begin{tabular}{|c|c|c|}
\hline 
\multicolumn{2}{|c|}{parameter} & value\tabularnewline
\hline 
\hline 
\multicolumn{2}{|c|}{$m_r$-reduce mass of $F_{2}^{-}$} & $9.5\,amu$\tabularnewline
\hline 
\multicolumn{2}{|c|}{$\Delta r$} & $0.0244\,bohr$\tabularnewline
\hline 
\multicolumn{2}{|c|}{$ r_{max}$} & $50\,bohr$\tabularnewline
\hline 
\multicolumn{2}{|c|}{Temperature} & $10\,K$\tabularnewline
\hline 
\multicolumn{2}{|c|}{Initial $J_{max}$ at $T=10\,K$} & $10$ \tabularnewline
\hline 
\multicolumn{2}{|c|}{$J_{max}$ at high intensity} & $20$ \tabularnewline
\hline 
\multicolumn{2}{|c|}{$\Delta t$} & $5\cdot10^{-2}\,fsec$\tabularnewline
\hline 
\multicolumn{2}{|c|}{$ N_t$ - number of time steps} & $7,500$\tabularnewline
\hline 
\multicolumn{2}{|c|}{Pulse width } & $30\,fsec$\tabularnewline
\hline 
\multicolumn{2}{|c|}{Central wavelength} & $350\,nm$\tabularnewline
\hline 
\multicolumn{2}{|c|}{Intensity} & $10^{10}\,\frac{W}{cm^{2}}\,,\,10^{13}\,\frac{W}{cm^{2}}$\tabularnewline
\hline 
\multicolumn{2}{|c|}{Electronic potentials and dipole momentum} & private communication with Prof. Robert Moszynski\tabularnewline
\hline 
\end{tabular}
\color{black}
\end{table}

The scalar product between the transition dipole operator, $\boldsymbol{ \mu }_{n,k}$, and the radiation field, \textbf{$\varepsilon$}, can generate  different transition schemes.
The specific scheme is influenced by the  states involved, symmetries and the field polarization. The details are elaborated in our previous work \cite{ezra2020dissociation}. The present study contains only two \color{black} $^1\Sigma$ singlet electronic state. Hence, under these conditions the nuclear rotational angular momentum operator merge to the total angular momentum of the molecule, $\hat{\textbf{R}}\left(\theta,\phi\right)=\hat{\textbf{J}}\left(\theta,\phi\right)$. In the absence of other angular momenta, and the Hund's case (a) \cite{richard1988angular}, the most common case and the one used here, the projection of $\boldsymbol{\hat{J}}\left(\theta,\phi\right)$ on the laboratory axis, $\hat{Z}$, is $m$ and on the internuclear axis is $\Omega$. \par

Similar to our previous work\cite{ezra2020dissociation}, we expand the angles' coordinate by the Wigner rotation matrices, $D^j_{m,\Omega}\left( \varphi,\theta \right)$.  
With this basis set, each state is described by $r,j,m$, and $\Omega$.  
The density matrix of the molecular state at time $t$ is:

\begin{equation}
    \hat{\rho}\left(r,r',\theta,\theta',\varphi,\varphi',n,n';t\right)=
   \sum_{\zeta,\zeta'}
    \ket{n}\bra{n'} \otimes
    a_{n,n',\zeta,\zeta'} \left(r,r';t\right) \otimes \ket{\zeta}\bra{\zeta'}
\end{equation}
where the summation is over all rotational states, $\zeta=\{j,m,\Omega\}$, thus $\ket{\zeta}$ is the wavefunction with the quantum numbers $\ket{j,m,\Omega}$, expanded by the $D^j_{m,\Omega}\left( \varphi,\theta \right)$ matrices. $a_{n,n',\zeta,\zeta'} \left(r,r';t\right)$ is the nuclear matrix coefficients at each electronic and rotational state. Since this work includes only $^1\Sigma$ states  $\Omega=0$, and the matrices are identical to the spherical harmonics $Y_{j,m}\left( \varphi,\theta \right)$. For linearly polarized external field, the coupling elements between the states due to the interaction with the field, are proportional to $ cos\left(\theta\right) \propto D^1_{0,0}\left( \varphi,\theta \right)$.\par

To mimic the experimental process, a thermal ensemble of molecules is subjected to a pulsed laser source. Within the temperatures considered, the electronic and vibrational degrees of freedom are in their ground state. The initial state is a Boltzmann distribution of several rotational components and can be written as:

\begin{eqnarray}
      \hat{\rho}\left(r,r',\theta,\theta',\varphi,\varphi',n,n';t=0\right)=\nonumber\\
        \sum_{\zeta_i} \mathtt{P}_B\left(j_i\right) a_{0,0,\zeta_i,\zeta_i}\left(r,r';t=0\right) \cdot \left|0\right>\left<0\right|   \otimes
     \mathsf{\hat{P}}_{0,j_i}(r) \otimes \left|\zeta_i\right>\left<\zeta_i\right| 
\end{eqnarray}

where $\mathsf{\hat{P}}_{0,j_i}(r)=\left|\Phi_{n=0,j_i}\left(r\right)\right>\left<\Phi_{n=0,j_i}\left(r\right)\right|$ is the projection onto the ground vibrational state, that conditioned by the rotational state, $j_i$. $ \mathtt{P}_B\left(j_i\right)={\frac{exp\left(-\beta E_{j_i}\right)}{\mathcal{Z}}}$ { are the Boltzmann coefficients}\color{black}, where $\mathcal{Z}=\sum_l \left(2l+1\right)exp\left(-\beta E_l\right)$ is the partition function. Note, that the rotational energy eigenvalues that satisfy  $\hat{H}_{rot} \left|\psi_{\zeta_i}\right> = E_{j_i} \left|\psi_{\zeta_i} \right>$, depend only on the total angular momentum $j$.

The time evolution of the state is carried out by a first principle propagation of the time-dependent Schr\"odinger equation: $\hat{\rho}\left(t\right)=\hat{U}\left(t\right)\hat{\rho}\left(t=0\right)\hat{U}^\dagger\left(t\right)$ where $i \hbar \frac{d}{dt} \hat{U}\left(t\right) = \hat H(t) \hat{U}\left(t\right)$. 
Observables are calculated by the data provided by the density operator, $\hat{\rho}(t)$ so that for any operator $\hat{A}$:
 \begin{equation}
     \left<\hat{A}\right>_t= tr\left(\hat{A}\rho\left(t\right)\right)  =tr\left(\hat{A}\hat{U}\left(t\right)\rho\left(t=0\right)\hat{U}^\dagger\left(t\right)\right)
 \end{equation}

Under the common experimental conditions for the  molecular beam considered here, the  typical incoherent processes such as collisions and spontaneous decay take place on the time scales of $t\ge 1\,nsec$. 
The dissociation under $30\, fsec$ laser pulse, is achieved by $\sim 100\,fsec$. The dynamics are therefore  coherent. 
Therefore, the expectation value can be evaluated, employing the basis where the initial density operator is diagonal $\left|\tilde{\psi}_{0,0,\zeta_i}\right>=\left|\psi^e_0\right> \otimes \left|\psi^\nu_0\right> \otimes \left|\zeta_i\right>$, giving by:

\begin{equation}\label{dinsityExVa}
     \left<\hat{A}\right>_t =
      \sum_{n,\nu,\zeta} \mathtt{P}_B\left(j_i\right) \left<\hat{U}^\dagger\left(t\right)\tilde{\psi}_{0,0,\zeta_i}\left(t=0\right)\right| \hat{A} \left|\hat{U}\left(t\right)\tilde{\psi}_{0,0,\zeta_i}\left(t=0\right)\right>
\end{equation}

Using equation (\ref{dinsityExVa}) we can decompose the expectation value into the expectation values of the individual components. 
Each component requires the solution for time-dependent Schr\"{o}dinger equation: $ i \frac{d}{dt} \psi = \hat H(t) \psi $. Numerical methods for the wave function propagation such as the Chebychev approximation scale as $N_S^2$ \cite{tal1984accurate, kosloff1994propagation, chen1999chebyshev,damari2016rotationalSarly}. 
Observables that are computed by equation (\ref{dinsityExVa}) will be used as a reference to the results that will be derived by the random phase method.

The size of Hilbert space of linear molecules, that is the number of occupied states $N_S$, scales linearly with the temperature.
The experiments considered on $F_2^-$ are carried out at an estimated temperature of $T=100K$. At this temperature, the Hilbert space size of the full ro-vibrational system is $N_S\sim10^6$. Therefore, computing the dynamics with an elevated temperature is exceptionally challenging.

The light induced coupling between the states further contributes to the complexity of the problem. At low intensity the transition can be approximated perturbatively which keeps the scale of interacting states to $N_S^2$, while at high intensities, strong fields, the process requires full and exact calculation, which increase the scale of the states to $N_S^3$. 

To overcome the challenge of the large number of states,  the random phase wavefunction method is used. 
An efficient representation of the state is done by the use of a limited ensemble of pure states, mimicking the initial thermal ensemble. The initial state of a single random phase wavefunction is written as:
\begin{equation}
    \ket{\Psi^k\left( r,\theta,\varphi;t=0\right)} = \sum_{\zeta_i} e^{i\chi_{\zeta_i}^k} \sqrt{\mathtt{P}_B\left(j_i\right)} \sqrt{\tilde{a}_{0,\zeta_i}\left(t=0\right)} \\
    \ket{0}\otimes\ket{\Phi_{0,\zeta_i} \left(r\right)}\otimes \ket{\zeta_i} \label{eq:inifitalState}
\end{equation}
where the phases $\chi_{\zeta_i}^k$ are chosen randomly, and ${\tilde{a}_{0,\zeta_i}\left(t=0\right)}$ are the coefficients corresponding to the diagonal elements of $a_{0,0,\zeta_i,\zeta_i}\left(r,r;t=0\right)$, and the angular part is introduced by $\ket{\zeta_i}$. The propagation of the wavefunction is done by {solving the time dependent Schr\"{o}dinger equation, $\ket{\Psi^k\left( r,\theta,\varphi;t\right)}=U\left(t\right)\ket{\Psi^k\left( r,\theta,\varphi;t=0\right)}$ (using the Chebyshev expansion\cite{chen1999chebyshev} and absorption boundary condition in additional auxiliary channels as fully described at our previous work\cite{ezra2020dissociation})}\color{black}. For large number of realization $k$, the set of random phase wavefunctions are complete basis of states:
\begin{equation}\label{eq:conDensityFunciton}
    \lim_{K\rightarrow\infty}\frac{1}{K}\sum_{k=1}^K \ket{\Psi^k\left( r,\theta,\varphi;t\right)}\bra{\Psi^k\left( r',\theta',\varphi';t\right)} = \hat{\rho}\left(r,r',\theta,\theta',\varphi,\varphi',n,n';t\right)
\end{equation}

The efficiency of the random phase wavefunction method increases with the size of Hilbert space, i.e., with the number of random phases, which increases with the temperature\cite{kallush2015orientation}.

The time dependent observables are computed for each random phase wavefunction and then averaged over $K$ realization. The error with respect to the exact solution, the reference, decreases with the number of realizations, $K$. The observables are calculated using the RPW as follows:
\begin{equation}\label{dinsityExVaRand}
     \left<\hat{A}\right>_t^K = 
      \frac{1}{K}\sum_{k=1}^K\,\,\,\sum_{n,\nu,\zeta} \bra{\Psi^k\left( r,\theta,\varphi;t\right)} \hat{A} \ket{\Psi^k\left( r,\theta,\varphi;t\right)}
\end{equation}

In the following section, we present the results calculated by using the random phase wavefunction method on the spatially resolved photo-dissociation. The quantum observables are then compared to numerically
exact computations\cite{ezra2020dissociation}, carried by propagation of all the initially occupied states and thermal summation of their expectation values.\color{black}
  
\section{Results}
\label{sec:results}

A  first principle \color{black} simulation has been designed to simulate the photo-dissociation process of a diatomic molecule. The RPW method has been employed to overcome the large Hilbert space challenge. Thus, the initial state was sampled from the Boltzmann thermal state by random phase wavefunctions, equation (\ref{eq:inifitalState}).
The photo-fragments outcome state is represented by the final density operator  composed from averaging the individual initial RPW states, Eq. (\ref{eq:conDensityFunciton}). 



It is worth mentioning, that for low intensities and temperatures, the small number of occupied states leads to a small number of random phases, which makes the method impractical. On the other hand, the dynamics for \color{black}high intensities and temperatures are impossible to compute exactly. We thus restrict the study of the system properties to low intensity and high temperature - $1\cdot10^{10}\left[\frac{W}{cm^2}\right], T=80\left[K\right]$ and high intensity and lower temperature- $1\cdot10^{13}\left[\frac{W}{cm^2}\right], T=50\left[K\right]$. In each calculation, the approximated RPW result was compared to the exact computation, carried by separately computing the dynamics for each initial state in the full Hilbert space.

\begin{figure}
    \centering
    \includegraphics[width=1.\textwidth]{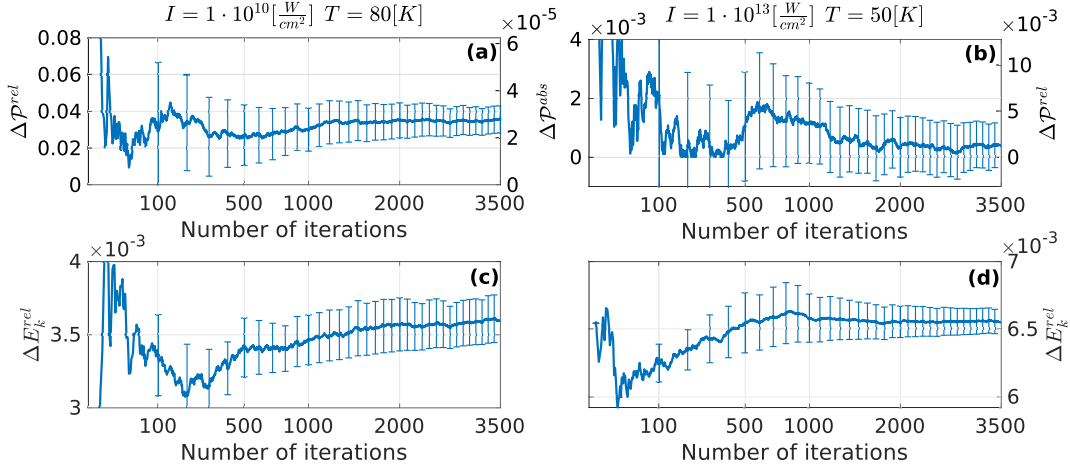}
    \caption{Observables errors as a function of the number of realizations: dissociation probability and kinetic energy.  Results for the low intensity, left panel-figures \textbf{a} and \textbf{c}, and high intensity, right panel-figures \textbf{b} and \textbf{d}. Figures \ref{fig:observablesErrorBar}\textbf{a}- \ref{fig:observablesErrorBar}\textbf{b} shows the error in the dissociation probability, $\mathcal{P}$. The error is displayed by two axis, the absolute difference between the probabilities, the inner axis. The outer axis display the relative error. Figures \ref{fig:observablesErrorBar}\textbf{c}- \ref{fig:observablesErrorBar}\textbf{d} present the relative error in the kinetic energy as a function of the number of realizations.}
    \label{fig:observablesErrorBar}
\end{figure}

First, the dissociation probability and the average kinetic energy of the fragments were calculated. The dissociation probability, $\mathcal{P}^k$, and the  kinetic energy, $\mathcal{E}^k$, for each realization $k$ from the final wavefunction are:

\begin{equation}\label{DissProbEq}
    \mathcal{P}^k =  \left<\Psi^k_{n\neq0}\left( r,\theta,\varphi;t=t_{final}\right)| 
     \Psi^k_{n\neq0}\left( r,\theta,\varphi;t=t_{final}\right)\right>
\end{equation}
\begin{equation}\label{EKProbEq}
    \mathcal{E}^k =  \left<\Psi^k_{n\neq0}\left( r,\theta,\varphi;t=t_{final}\right)\left|\hat{P}_r^2\right|
     \Psi^k_{n\neq0}\left( r,\theta,\varphi;t=t_{final}\right)\right>
\end{equation}
\color{black}
The exact dissociation probability, calculated according to equation (\ref{DissProbEq}), is $0.313$  at the high intensity and $7.81\cdot10^{-4}$ at the low intensity. 
Figure \ref{fig:observablesErrorBar} presents the deviation of the approximate computation from the exact one, for the dissociation probability $\mathcal{P}$, and the fragments' average kinetic energy $E_k$.

\color{black}
The error bars correspond to the standard deviation of the results from the mean value at each given number of realizations. For any observable $A$ this is given by: $\frac{1}{K}\sqrt{\sum_{k=1}^K(<A>_k-<A>_K)^2}$, where $<A>_K$ is the averaged value of $A$ after $K$ realizations and $<A>_k$ is the value for a given realization $k$\color{black}. The convergence of the deviation with the number of realizations is visible.

Figure \ref{fig:observablesErrorBar}\textbf{a} - \ref{fig:observablesErrorBar}\textbf{b} displays the error of the dissociation probability. Two scales are displayed: the relative error $\Delta{\cal{P}}_{rel} = {{\left|{\cal{P}}_{RPW}-{\cal{P}}\right|}\over{{\cal{P}}}}$ in the outer axis and absolute error $\Delta{\cal{P}}_{abs} = \left|{\cal{P}}_{RPW}-{\cal{P}}\right|$ in the inner axis. As expected, at the high intensity the large overall ${\cal P}$ leads to an absolute error much larger compared to the low intensity. The relative error of this case is, however, significantly lower than at the low intensity. Note that a comparison to the experimental output reveals that, at least for commonly used methods, the resolution for the low intensity regime is way beyond the experimental. Absolute experimental values for the dissociation probability for the higher intensity regime are even harder to be determined. 

The kinetic energy of the photo-fragments is used experimentally to track the asymptote of the electronic state from which the dissociation takes place. 
Panels (c) and (d) in Figure \ref{fig:observablesErrorBar} present the average kinetic energy relative error. The error at both intensities is already lower than $1\%$ after $200$ realizations.

\begin{figure}
    \centering
    \hspace*{-3cm} 
    \includegraphics[width=1.5\textwidth]{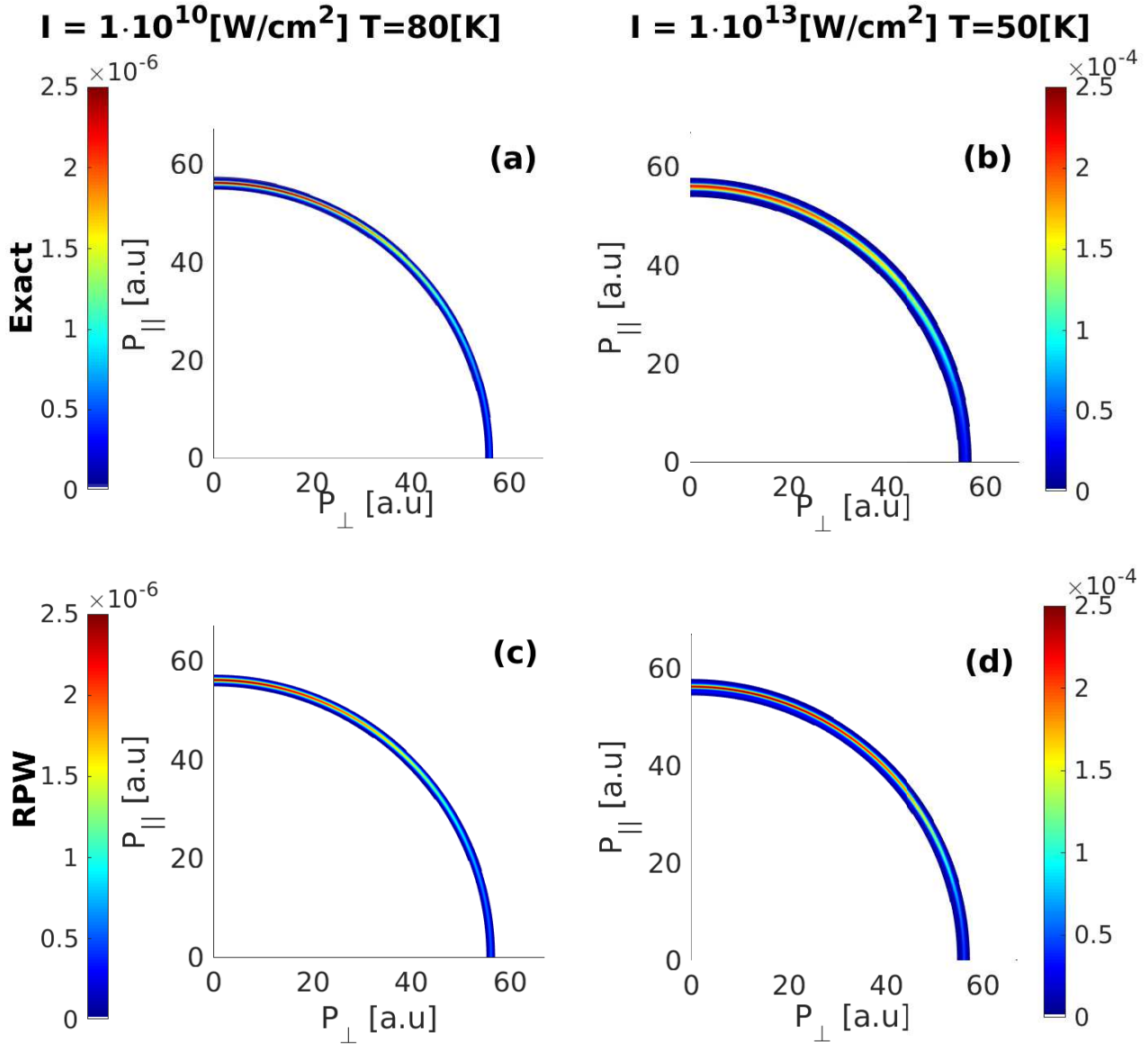}
    
    \caption{Angular distribution of the outgoing momentum, $P$\color{black}, shown as a density plot. Exact (Upper panels a-b) and RPW method after 3500 realizations (Bottom panels c-d). Left and Right panels are for low and high intensities, respectively. }
    \label{fig:dis2Cas}
\end{figure}

Figure \ref{fig:dis2Cas} presents the angular distribution of the momentum fragments calculated by the direct method, figures \ref{fig:dis2Cas}a-\ref{fig:dis2Cas}b, and the RPW method, figures \ref{fig:dis2Cas}c-\ref{fig:dis2Cas}d. 
The angular distribution of the out going momentum, $P$, is calculated as follows: 

\begin{equation}\label{eq:AngularMomentumDistribtion}
    \mathcal{D}^k \left( P,\theta \right) = \\
    \int d\phi \sum_{m,\Omega,n>0} \left| \sum_j \sum_{\zeta_i} e^{i\chi_{\zeta_i}^k} \sqrt{\mathtt{P}_B\left(j_i\right)} \sqrt{\tilde{a}_{n,0,\zeta,\zeta_i}\left(t=t_{final}\right)} \cdot  D^{j}_{m,\Omega}\left(\theta,\phi\right)  \right|^2
\end{equation}
{Where $\tilde{a}_{n,0,\zeta,\zeta_i}\left(t=t_{final}\right)$ are the coefficients in momentum representation corresponding to the ${a}_{n,0,\zeta,\zeta_i}\left(t=t_{final}\right)$ coefficients. }\color{black}
Note that the summation over $j$ is a quantum summation that conserves the phases while the summations over $m$ and $\Omega$ are classical. The different is an outcome of the dynamics, which generates coherences only between different values of $j$ with the same values of $m$ and $\Omega$. 
\color{black} 

The two distributions are qualitatively similar, with the same central momentum peaks and parallel orientation. However, changes in small details could be observed, especially  when comparing the distributions at high intensities panels of figure \ref{fig:dis2Cas}d and figure \ref{fig:dis2Cas}b.

\begin{figure}
    \centering
    \includegraphics[width=1\textwidth]{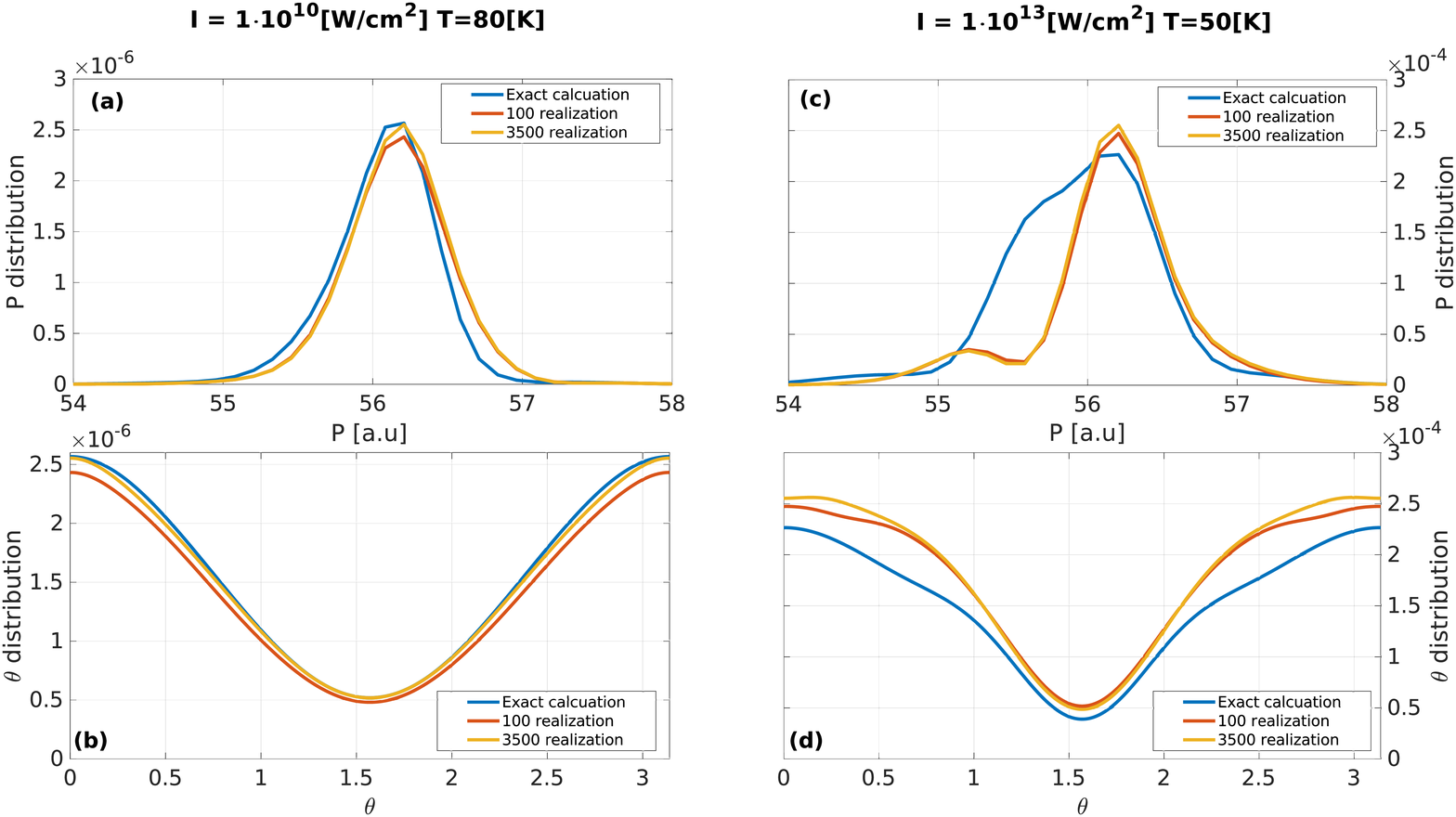}
    \caption{\textbf{Upper panels}: The outgoing momentum distribution as a function of the momentum in the parallel direction $P_\parallel$. The exact calculation is compared to results obtained with different number of realizations. The panels  correspond to low laser intensity figure\ref{fig:PAndThetaDist2Cas}a, and high laser intensity, figure\ref{fig:PAndThetaDist2Cas}c. \textbf{Lower panels}: The outgoing momentum distribution as a function of the angle $\theta$ for the exact calculation and for different number of realizations. Low laser intensity, figure\ref{fig:PAndThetaDist2Cas}b, and high laser intensity, figure\ref{fig:PAndThetaDist2Cas}d.
    The figures sample a cut in the corresponding distribution in figure \ref{fig:dis2Cas}, see text. Mind that while the angular axis span the whole range of values for $\theta$, the presented result for the $P$ axis is about $5\%$ of the total grid, thus its resolution is significantly lower. }
    \label{fig:PAndThetaDist2Cas}
\end{figure}

A more detailed analysis is obtained by comparing separately the directions in the azimuth and the absolute value of the momentum. The upper panels in figure \ref{fig:PAndThetaDist2Cas} presents the distribution as a function of the momentum for $P_\perp=0$. Note, that only a very narrow range of values from the full momentum grid are plotted. The lower panel at figure \ref{fig:PAndThetaDist2Cas} presents the distribution as a function of the angle $\theta$, taken at the most probable momentum. At low intensity, the convergence of both distributions to the exact distribution is fast, while at high laser intensity, the distribution has not yet converged to the  exact calculation even after $3500$ realizations. This discrepancy is the result of significant interference at high intensity and high coherence. Specifically we see the sensitivity to interference on the momentum and angular distribution. Atop of this, one should beware of the fact that the grid resolution at the $P$ axis is poor, due to the very narrow width of the distribution in the momentum space. 
 
The distance between the distributions as a function of the number of realizations is calculated by the Wootters metric\cite{wootters1981statistical}. The distance is calculated for the momentum distribution, $d_W^p$, and at the angular distribution, $d_W^\theta$, defined by the following equations:

\begin{equation}\label{eq:dwooP}
    d_W^p\left(K\right)=\frac{\pi}{2}cos^{-1}\left(\int P^2 dP\sqrt{\mathcal{D}_{rand}^K(P,\theta=0)}\cdot\sqrt{\mathcal{D}_{exact}(P,\theta=0)}\right)
\end{equation}

\begin{equation}\label{eq:dwooTheta}
    d_W^\theta\left(K\right)=\frac{\pi}{2}cos^{-1}\left(\int sin(\theta)d\theta\sqrt{\mathcal{D}_{rand}^K(P=P_{mp},\theta)}\cdot\sqrt{\mathcal{D}_{exact}(P=P_{mp},\theta)}\right)
\end{equation}

\begin{figure}
    \centering
    \includegraphics[width=1.\textwidth]{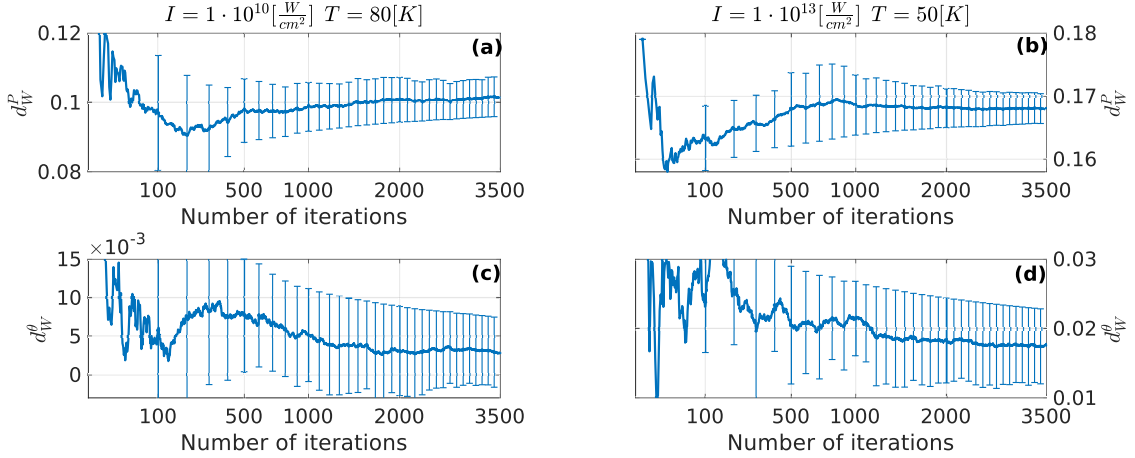}
    \caption{Distributions' distance as a function of the number of realizations. Arranged similarly to fig. \ref{fig:PAndThetaDist2Cas}. Figures \ref{fig:ErrorFig2Int}\textbf{a}- \ref{fig:ErrorFig2Int}\textbf{b}
    displayed the distance between the momentum distributions comparing the random phase method to the exact calculation, $d_W^p$,  equation \ref{eq:dwooP}, while figures \ref{fig:ErrorFig2Int}\textbf{c} \ref{fig:ErrorFig2Int}\textbf{d}, preset the distance of the angular distributions, $d_W^\theta$,   equation \ref{eq:dwooTheta}.}
    \label{fig:ErrorFig2Int}
\end{figure}

We choose to calculate the distribution at the momentum direction, $d_W^p$, and the angular direction,  $d_W^\theta$, separately. For each calculation we set one of the direction at the most probable value, $\theta =0$ and $p=p_{mp}$.
\color{black}
The distance was normalized so that its maximal value is $1$ for non overlapping distributions. 
The distances between the distributions as a function of the number of realizations are present in figure \ref{fig:ErrorFig2Int}.

The convergence of the distribution is analyzed by the distance between the distributions at different laser intensities. An inspection of the distances in figure \ref{fig:ErrorFig2Int}, reveals that the distribution at low intensity, both in angular and momentum, are closer to the exact distributions. Furthermore, in both intensities, the momentum distribution is closer to the exact solution than the angular distribution. For an extremely narrow distribution like in out case, the distribution distance is sensitive to small differences and account for high levels of coherences. In contrast, observables which are more classical and contain lower level of off diagonal elements in the energy representation, such as the average kinetic energy and dissociation probability, are less sensitive to coherences, and are much easier to capture with the random phase method.

\section{Discussion $\&$ Conclusions}

In this work, a computational method for the  first principle \color{black} simulation of strong field photo-dissociation and other time domain spectroscopy was developed.  Simulations based on an exact solution of the time-dependent Schrödinger equation are expensive. Typical computational complexity  scales as $N_S^3$, where $N_S$ is the number of state that are involved in the process. Simulations of diatomic photo-dissociation process at typical experimental conditions require
a vast Hilbert space, of the order $10^6$. 

To overcome this challenge, the RPW method has been employed.
The phases of the initial states are randomly chosen and  thermal averaged observables are obtained. The observables are converged to the directly calculated outcome after $K$ realizations, where $K<<N_S$. Previous studies on RPW \cite{gelman2003simulating,kallush2015orientation,hu2021investigation} showed that the observable convergence scales as $\sim \frac{1}{\sqrt{K}} \cdot \frac{1}{\sqrt{N_S}}$. In this study, the required number of $K$ realizations to converge an observable to the experimental accuracy is elucidated.

As RPW method relay on randomness of the relative phase between the states, it is unlikely that processes which are driven by high order coherent interactions, e.g., multi-photon transitions, would be well approximated by the method. This type of processes are highly coherent and employ throughout the dynamics non diagonal elements of the density matrix, i.e., coherences.  
\color{black}

The dissociation probability, $\mathcal{P}$, is an example of observable in which the only contribution of the observable comes from diagonal elements of the density matrix in the energy representation. Hence, coherent interferences, have low impact on such observable. \color{black} Another case is the average kinetic energy, $\Delta E^{rel}_k$, an observable that is less affected by the different interference pathways in the dynamics. The average kinetic energy, at the two intensities, converges to the exact solution up to the experimental error of about $0.02eV$ after only $100$ realization, which means acceleration of the calculation by a factor of $\sim 50$ compared to the direct method. 
An explanation can be attributed to the property of the operators that correspond to these observables that are diagonal in the energy representation. Thus, they are less dependent on coherences and interferences. 

Another observable that was explored is the momentum spatial distribution of the dissociated fragments. These distributions provide a visual approach to study the convergence. Our main result concerning this quantity is that its dominate features are accurately recovered by the method, while fine details might be difficult to retrieve. In addition, the convergence is better for the angular variable $\theta$ than for the magnitude of the momentum, due to the very different width of the distribution in these two directions. We have to remark here, that the common use of this tool in the experimental context is usually more quantitative. Its main use is to assign the different fragments to their channel within the different electronic curves. Therefore, our examination of the convergence here, and the finding that the exact distribution is difficult to recover affects the applicability of the method only marginally.   

The observed error leads us to conclude that different observables converge differently. Each observable is sensitive to different properties of the photo-fragments state. The dynamics of the dissociation process includes interferences. Adding random phases changes the interference pattern, leading to different outcomes. At high intensity, the process contains multi-photon transitions that induce rare events and interferences that lead to special states. Thus, adding a random phase leads to different events and interferences, and to different final states. 

The observables convergence is subject to the coherence and interference during the process. We note hereby, that the choice of the wavefunction basis, might be another factor that effects the observable convergence. However, upon changing the basis from the Wigner $D$ matrices to a direct angular grid representation basis, we did not observe any significant shift in the convergence rate. 

In conclusion, we constructed a first principle \color{black} model for time-domain spectroscopy, for calculating experimental properties. The dynamical processes are computationally complex due to the large Hilbert space; thus we used the RPW method to reduce the complexity. 
We studied the convergence properties of the product observables at two laser intensities. The same method would apply to the dynamical hole left in the reactants \cite{banin1994impulsive}.
We conclude that the convergence depended on the sensitivity of the observable to the dynamic interferences that are taking place in the process.
Thus, at high laser intensity, where the interference leads to the specific final states, the convergence of sensitive observable is slower.

\section*{Acknowledgements}
The authors thanks Daniel Strasser, Robert Moszynski, 
Iwona Majewska, Aviv Aroch. This research was supported by the Israel Science Foundation (Grants No. 510/17, 526/21).

\bibliography{mybibliography}

\end{document}